\documentclass[a4paper]{article}
\usepackage[right=2.5cm,left=2.5cm,top=3.5cm,bottom=3.5cm,headsep=0cm,footskip=0.5cm]{geometry}
\usepackage{graphicx}
\usepackage{amssymb}
\usepackage{slashed}
\usepackage{color}

\title{\textbf{Reflections on Cosmology: \\
an Outsider's Point of View}}
\author{Adolfo Toloza$^{1,2}$ and Jorge Zanelli$^{1,3}$\\
$^1$  {\small \textit{Centro de Estudios Cient\'{\i}ficos, Arturo Prat 514, Valdivia, Chile.}} \\
$^2$ {\small \textit{Instituto de F\'{\i}sica, P. Universidad Cat\'olica de Valpara\'{\i}so,  Casilla 4059, Valpara\'{\i}so, Chile.}} \\ 
$^3$ {\small \textit{Universidad Andr\'es Bello,  Av. Rep\'ublica 440, Santiago, Chile.}}}
\date{}

\begin{document}
\maketitle

\begin{abstract}
Some of the assumptions of cosmology, as based on the simplest version of General Relativity, are discussed. It is argued that by slight modifications of standard gravitation theory, our notion of the sources of gravity --the right hand side of Einstein's equations--, could be something radically different from what is usually expected. One example is exhibited to prove the point, and some consequences are discussed. 
\end{abstract}

\section{Introduction}  
Almost a hundred years have passed since Einstein's formulation of General Relativity (GR). According to this theory, gravitation is a manifestation of geometry: the presence of matter and energy curves spacetime and, consequently, the trajectories of particles in spacetime are bent in the presence of matter or energy. In spite of the remarkable success of GR as a predictive and accurate model for the cosmos, the outstanding open problem that has lingered for the past eighty years has been how to reconcile GR with quantum mechanics. The fundamental microscopic nature of the spacetime manifold has been a mystery that has prompted the development of new areas of research at the interface between high energy and gravitation, like string theory and loop quantum gravity. Unfortunately --or fortunately, depending on one's point o view--, the lack of experimental evidence of quantum gravitational effects has turned the study of quantum gravity into a rather academic mathematical exercise.

On the other hand, the large scale behaviour of the cosmos has been rich in counter intuitive observations --like the accelerated expansion of the universe-- and puzzling conclusions --like the existence of exotic fluids ($p<-1/3 \rho$)--, or strange matter fields driving inflation. Moreover, even at the galactic scale, GR is not sufficient to describe all observed phenomena, like the star velocity distribution. The standard cosmology, based on the simplest assumptions of GR, seems to lead to exotic models of mysterious origin: dark matter, dark energy, inflation.

The difficulty of GR to make room for quantum phenomena is excusable: it was not meant to be appropriate to describe the microcosm; had it turned out to be adequate in this realm, it would have been an unexpected and undeserved triumph. In contrast, GR's inability to give a natural, coherent description of the universe at large, is a major failure. It could be argued, that at the beginning of the twentieth century, the known universe for which the theory was tailored, extended basically to our galaxy and astronomy was essentially restricted to a few centuries-light around the solar system. One hundred years ago, nobody knew whether anything existed beyond our galaxy and the current reach of telescopes was certainly beyond physicist's wildest expectations. 

At any rate, the cosmological riddles seem to be a challenge for GR within reach of our present understanding: it does not need quantum postulates and it is experimentally verifiable. It requires, however, to think again about what we mean by spacetime geometry and how it relates to matter and energy. 

\subsection{What (we believe) we know}  

The current view, shared by most physicists --and many non-physicists as well--, is that the universe started with a big splash about fourteen billion years ago, evolving according to the so-called \textit{standard big-bang cosmology}. The main ingredients in this recipe are found in GR, including \\
$\bullet$ Principle of equivalence: spacetime is invariant under local Lorentz transformations. \\
$\bullet$ Einstein's equations (\textbf{EE}), $G_{\mu\nu} =T_{\mu\nu}$: matter and energy curve spacetime.\\ 
$\bullet$ Friedmann-Robertson-Walker-Lemaitre (FRWL) cosmology: simplest homogeneous, isotropic, expanding solution of EE.

The evidence for the validity of this model is inferred from astronomical observations which seem consistent with the following assertions:\\
$\bullet$ $13.7 \times 10^9$ years ago, the universe was hot, fairly homogeneous and isotropic\\
$\bullet$ The spatial section of the universe is quite flat at large scales\\
$\bullet$ The universe is very near the critical density\\
$\bullet$ The universe has been expanding at an increasing rate (accelerated expansion) .

There is a narrative that comes with this picture that leads to the main puzzles of current cosmology. This includes concepts such as dark matter, inflation and dark energy, which arose from the frustrated efforts to account for the observed motion of stars in galaxies, and the remarkable coincidence of parameters, that produces a universe of extremely flat and homogeneous spatial sections, in accelerated expansion.

Most physicists look at coincidences with suspicion and tend to prefer a causal explanation, even if far-fetched. For want of better explanations, dark and inflationary ideas seem currently acceptable for most of the astrophysics community. Just to get a flavour of the type of understanding, let us quote a typical web page where these ideas are presented to a broad audience \cite{NASA}: 

``[On the other hand,]...measurements of the cosmic microwave background indicate that the Universe has a flat geometry on large scales. Because there is not enough matter in the Universe --either ordinary or dark-- to produce this flatness, the difference must be attributed to a "dark energy". This same dark energy causes the acceleration of the expansion of the Universe.... The name "dark energy" refers to the fact that some kind of "stuff" must fill the vast reaches of mostly empty space in the Universe in order to be able to make space accelerate in its expansion." 

The theoretical challenge is to account for these observations with a dynamical model to explain these effects: a modification of the Einstein-Hilbert action, or the equations of state for the matter-energy content, new matter fields, etc. The physics community is a conservative lot, and a modification of the standard lore is not the first option. However, the challenges brought about by the astrophysical evidence suggests the need for revision of some of the original assumptions. On the other hand, the cosmic riddles seem to be within reach of our present understanding: they do not need quantum postulates and seem observationally verifiable. It requires, however, to go back to the drawing board and think again about what we mean by spacetime geometry and how it relates to matter and energy at a macroscopic scale. 

\section{Accepted wisdom}  

Ever since Einstein, the spacetime geometry has been assumed to be given by a pseudo-Riemannian torsion-free manifold, described by the metric through Einstein's equations, 
\begin{equation}
R^{\mu}_{\, \, \nu} -\frac{1}{2} R\delta^{\mu}_{\, \, \nu}  - \Lambda \delta^{\mu}_{\, \, \nu} =8\pi GT^{\mu}_{\, \, \nu}\, .
\end{equation} 
The evidence of the spacetime curvature is obtained by observing the geodesics of celestial objects and light. We believe to be able to clearly distinguish between geometry and matter-energy: The geometry is described by the metric whose curvature defines the left hand side (LHS), while the matter-energy content of spacetime defines the right hand side (RHS) of the equations. This framework is a natural extension of the experience in mechanics and electrodynamics, where the sources (RHS) determine the evolution of the relevant dynamical variables --positions and field configurations-- in the LHS. 

This point of view is so ingrained in or way of thinking about spacetime that we infer the matter content of the universe from the motion of particles. If the geodesics are those of a flat spacetime, we assume there is no matter anywhere. Conversely, if the geodesics are not straight lines, we infer that there must be matter somewhere, even if invisible. This, however, is not true and counterexamples of both statements are known. In the first case, it has been shown that conformally coupled matter does not always curve spacetime \cite{AMTZ}. 

As we shall see next, spacetime curvature is not necessarily due to the presence of matter sources; the source of curvature might be geometry itself.

\section{Other options} 

In Einstein's description of gravity, the relevant dynamical variable is the metric, while the torsion is assumed to vanish identically. The torsion-free condition is also justified in a broader sense. If the metric and the affine connection are assumed to be dynamically independent and independently varied in the action, then in four-dimensions and without fermionic matter, the field equations still force the torsion to vanish. Thus $T^a\equiv 0$ is not a restriction in the classical theory but a consequence of the equations of motion, which means that imposing this condition is not really necessary.

Still, even if the condition $T^a\equiv 0$ passes the experimental tests, it is theoretically unsatisfactory to impose it, as setting would be to impose the magnetic field identically zero to describe the interaction between a charged particle and electrostatic fields: within the experimental errors this might be o.k. and the statement $\textbf{B}=0$ could be a solution of Maxwell's equations in vacuum, but it is at best true in a particular reference frame, and certainly false in the the proper frame of a moving charge.

Instead of assuming the presence of exotic forms of invisible (dark) energy-matter (RHS), a slight modification in the dynamics of geometry (LHS) can be assumed. Numerous modifications of this sort to the Einsten-Hilbert action have been proposed. In some cases, extra terms are included in the Action ($f(R)$, $f(T)$ \& $f(\mathcal{G})$ theories), but this brings about severe difficulties, as the higher order derivatives in the equations and the arbitrariness in the form of $f$ (see for instance \cite{de-Felice,Capozziello,Nojiri,Ferraro}).

If $T^a$ is not assumed to vanish a priori, spinning particles would reveal the presence of torsion because they couple to $T^a$, making their trajectories to deviate from the geodesics for spinless matter. These effects may not be significant at the current experimental accuracy, or it might be too difficulty to compare trajectories of similar particles with different spin, in order to separate the effects of curvature and torsion in a spacetime region. 

Another aspect of torsion is that it could also act as an indirect source of curvature, and this may have important cosmological consequences. Here we examine an example of this effect.

\section{First order formalism}  

In the first order formalism of gravitation, the fundamental fields are the vielbein ($e^a=e^a{}_{\mu} dx^\mu$) and the Lorentz (``spin") connection ($\omega^{ab}=\omega^{ab}{}_{\mu} dx^\mu$). These 1-forms correspond to two independent aspects of the geometry --the metric and the affine structures, respectively--, and hence are considered as dynamically independent in the action (see, e.g., \cite{zReview}). In this formulation, the four-dimensional Einstein-Hilbert action with cosmologial constant reads
\begin{equation}
I_{GR}[e,\omega] = \int_M \epsilon_{abcd} \Big(R^{ab}-\frac{\Lambda}{6} e^a e^b\Big) e^c e^d 
\end{equation}
Varying with respect to the vielbein and the Lorentz connection gives the Einstein equations and the torsion-free condition
\begin{equation}
\epsilon_{abcd} \Big(R^{ab}-\frac{\Lambda}{3}e^a e^b\Big)e^c=0, \quad \epsilon_{abcd} T^ae^b=0.
\end{equation}

In general, the spin connection can be conveniently separated into a torsionless part and the \textit{contorsion}, 
$\omega^{ab}=\bar{\omega}^{ab}+\kappa^{ab}$, where
\begin{equation}
0 = d e^a + \bar{\omega}^a_{\,\,b} \wedge e^b\rightarrow \bar{\omega}^a_{\,\,b \mu}=e^a_\nu (\nabla_{\mu} E^\nu_b), \label{0-torsion}
\end{equation}
and the contorsion is proportional to the torsion,
\begin{equation}
\label{Torsion-contorsion} T^a = \kappa^a_{\,\,b} \wedge e^b\, .
\end{equation}
In (\ref{0-torsion}), $E^\nu_a$ is the inverse vielbein and $\nabla$ is the covariant derivative in the Christoffel connection. In this scheme, the Lorentz curvature, $R^{ab} = d \omega^{ab} + \omega^a_{\,\,c}\wedge \omega^{cb}$, separates into de usual torsion-free part of Riemannian geometry, and a remaining part that depends on the torsion,
\begin{equation}
\label{Curvature} R^{ab} = \bar{R}^{ab} + \bar{D} \kappa^{ab} + \kappa^a_{\,\,c} \wedge \kappa^{cb}
\end{equation}
Where $\bar{R}^{ab} = d \bar{\omega}^{ab} + \bar{\omega}^a_{\,\,c} \wedge \bar{\omega}^{cb}$ is the Riemannian curvature and $\bar{D}$ is the covariant derivative for the torsionless conection $\bar{\omega}^{ab}$. If torsion is nonzero, the remaining terms would contribute to the field equations as a source for $\bar{R}^{ab}$. 

\subsection{A Minimal Modification}   

The next term in the Lovelock series after the Einstein-Hilbert Lagrangian is the Gauss-Bonnet density\footnote{From now on, wedge product between forms is always assumed.} --the Euler form, $\epsilon_{abcd} R^{ab} R^{cd}$--, which defines a topological invariant in four dimensions. Its addition does not modify the Einstein equations in the bulk, only affecting the boundary conditions and renormalizes the global charges \cite{ACOTZ}.\footnote{A similar situation occurs if other topological invariants like the Pontryagin form, $R^a{}_b R^b{}_a$, or Nieh-Yan form, $T^aT_a-R^{ab} e_a e_b$, are included.}

A non-trivial modification would be to include a topological density 4-form but with a nonconstant coefficient $\phi(x)$, as
\begin{equation}
I[e,\omega, \phi] = I_{GR}[e,\omega] + \int_{\bar{M}} \phi \epsilon_{abcd} R^{ab} R^{cd}
\end{equation}

If $\phi$ is a slowly varying function at cosmic scales, for all practical purposes this modification could be locally ignored.  However, the presence of $\phi$ has two other consequences that must be considered. First, there is one more field equation coming from the variation with respect to $\phi$. Second, the first order formulation is no longer equivalent to the standard metric approach of GR, because the torsion can no longer be set to vanish identically.\footnote{A similar modification proposed in \cite{Jackiw-Pi} considered the Pontryagin density instead of the Euler form, and used the second order formalism, so that torsion was implicitly assumed to vanish. Other approaches that also considered the Pontryagin form but in the first order formalism, \cite{Alexander1,Alexander2,Botta-Cantcheff,Cambiaso,Umit-Ertem}. However, no one seems to have considered the Euler form, which is more naturally obtained by dimensional reduction from an action in the Einstein-Lovelock family.}

Varying with respect to $e^a$, $\omega^{ab}$ and $\phi$, yields the following field equations
\begin{eqnarray}
\label{Einstein} \delta e^d &:& \epsilon_{abcd} \Big(R^{ab} - \frac{\Lambda}{3} e^a e^b \Big) e^c = 0 \\
\label{Torsion} \delta \omega^{ab} &:& \epsilon_{abcd} (T^a e^b + d\phi R^{ab}) = 0 \\
\label{Euler} \delta \phi &:& \epsilon_{abcd} R^{ab} R^{cd} = 0.
\end{eqnarray}

Equation (\ref{Einstein}) has the same form as the Einstein equations with the difference that since now Eq. (\ref{Torsion}) implies that torsion no longer vanishes, $R^{ab}$ is not the usual metric in a Riemannian curvature of GR. Using (\ref{Curvature}) in (\ref{Einstein}) one obtains
\begin{equation}
\epsilon_{abcd} \Big(\bar{R}^{ab} -\frac{\Lambda}{3} e^a e^b \Big) e^c = -\epsilon_{abcd} (\bar{D} \kappa^{ab} + \kappa^a_{\,\,s} \kappa^{sb}) e^c\, , \label{newEE}
\end{equation}
where the torsion terms have been taken to the RHS to emphasize that they can act like matter sources.

Note that $\phi$ need not be a fundamental field. Its origin is not specified  and there might be additional kinetic and potential terms in the action depending on it. The field $\phi$ could be a remnant from a dimensional compactification of a theory appropriate for higher energies, or a condensate of fermion fields, or some effective field resulting from a more fundamental dynamical ingredients.  Hence, although Eq. (\ref{Euler}) may not be an accurate account of the real situation, we take it as it is in order to illustrate what could be the consequences of a rather minimal modification.

On the other hand, Bianchi identities impose some consistency conditions on the equations that must be held by the solutions. Taking the exterior covariant derivative of (\ref{Einstein}) and (\ref{Torsion}) gives, respectively
\begin{eqnarray}
\label{CCE}  && \epsilon_{abcd} (R^{ab} - \Lambda e^a e^b) T^c = 0 \\
\label{CCT}  && \epsilon_{abcd} {R^a}_p e^p e^b = 0.
\end{eqnarray}

\section{Effect on cosmology}   

Consider a spacetime foliated by a family of isotropic and homogeneous, flat three-dimensional spatial slices (Bianchi type I)\footnote{We assume this geometry not because we expect the universe to be flat, but as an example to illustrate the point. The  general case $k \neq 0$ will be discussed elsewhere.}, as described by the standard FRWL cosmology with $k=0$.  The metric, 
\begin{equation}
ds^2 = -dt^2 + a^2(t)\Big[dr^2+r^2(d\theta^2 + \sin^2\theta d\varphi ^2)\Big],
\end{equation}
admits global Killing vectors $\pounds_\xi g_{\mu \nu} = 0$, associated to spatial translations ($\xi_{(i)}= \partial_i$) and rotations ($\xi_{(ij)}=x_i \partial_j - x_j \partial_i$).

Assuming torsion and the scalar field to have the same isometries as the background spacetime, we demand $\pounds_\xi {T^{\alpha}}_{\mu \nu} = 0$ and  $\pounds_\xi \phi = 0$. It is straightforward to see that the torsion and scalar field can only depend on time, and the only nonvanishing components of ${T^{\alpha}}_{\mu \nu}$ are\footnote{Here $i,j,k=\{1,2,3\}$ are spacetime and $I,J,K=\{1,2,3\}$ are Lorentz indices.} \cite{Xi-chen Ao-Xin-zhou},\cite{Andre Tilquin-Thomas Schucker}
\begin{equation}
{T^i}_{j t} \sim \delta^i_j \,\,\, ; \,\,\, {T^i}_{jk} \sim \epsilon^i{}_{jk}
\end{equation}
With the convenient choice of the vielbein $e^0=dt$ and $e^1 = a(t) dx$, $e^2 = a(t) dy $, $e^3 = a(t) dz$, the torsion 2-form can be written as
\begin{eqnarray}
\label{T0} T^0 &=& 0\\
\label{Ti} T^I &=& h(t) e^I e^0 + f(t) \epsilon^I{}_{JK} e^J e^K \, ,
\end{eqnarray}
where $h(t)$ and $f(t)$ are unspecified functions of time. From (\ref{Torsion-contorsion}) we identify the contorsion one-form to be $\kappa^{0I} = h(t) e^I$ and $\kappa^{IJ} = -f(t) \epsilon^{IJ}{}_K e^K$. On the other hand, the nonvanishing components of the torsionless part of the spin connection are $\bar\omega^{0I} = H e^I$, where $H = \dot{a}/a$  is the Hubble function, and therefore the spin connection reads
\begin{eqnarray}
\label{w0i} \omega^{0I} &=& (H+h) e^I\\
\label{wij} \omega^{IJ} &=& -f \epsilon^{IJ}{}_K e^K\, .
\end{eqnarray}
The Lorentz curvature $R^{ab} = d \omega^{ab} + {\omega^a}_c \omega^{cb}$ takes the form
\begin{eqnarray}
\label{R0i} R^{0I} &=& [(\dot{H} + \dot{h}) + H (H + h)] e^0 e^I + f (H + h) \epsilon^I{}_{JK} e^J e^K \\
\label{Rij} R^{IJ} &=& [(H + h)^2 - f^2] e^I e^J + (\dot{f} + H f) \epsilon^{IJ}{}_K e^K e^0
\end{eqnarray}
Replacing (\ref{T0}, \ref{Ti}, \ref{R0i}, \ref{Rij}) into the field equations (\ref{Einstein}-\ref{Euler}), one obtains
\begin{eqnarray}
\label{1} &&U^2 - f^2 - \frac{\Lambda}{3} = 0\\
\label{2} &&2(\dot{U} + H U) + U^2 - f^2 - \Lambda = 0\\
\label{3} &&-h + \dot{\phi} (U^2 - f^2) = 0\\
\label{4} &&\Big(-\frac{1}{2} + \dot{\phi} U \Big) f = 0\\
\label{5} &&(\dot{U} + H U)(U^2 - f^2) - 2 f U(\dot{f} + H f) = 0
\end{eqnarray}
where $U = H + h$. Equations (\ref{1}, \ref{2}), that come from (\ref{Einstein}), correspond to the modified Friedmann equations. If expressed in the standard form: $H^2 - \frac{\Lambda}{3} = \frac{\kappa^2}{3} \rho_{eff}$ and $2\dot{H} + 3H^2 - \Lambda = -\kappa^2 p_{eff}$ (where $\kappa^2 = 8 \pi G$), we recognize the effective density and pressure induced by torsion to be 
\begin{eqnarray} \label{rho}
\rho_{eff} &=&-\frac{3}{\kappa^2}(2 H h + h^2 - f^2) , \\
\label{p}
p_{eff} &=& \frac{1}{\kappa^2}(2\dot{h} + 4 H h + h^2 - f^2),
\end{eqnarray}  wich satisfy the continuity equation 
\begin{equation}
\dot{\rho}_{eff} + 3 H (\rho_{eff} + p_{eff}) = 0 ,
\end{equation}
as can be checked by direct application of (\ref{3}), (\ref{4}) and (\ref{5}). This result is related to the fact that our anzats satisfies the consistency conditions (\ref{CCE}, \ref{CCT}) which in turn, are consequences of the Bianchi identity.

\begin{figure}
\includegraphics[width=0.5\textwidth]{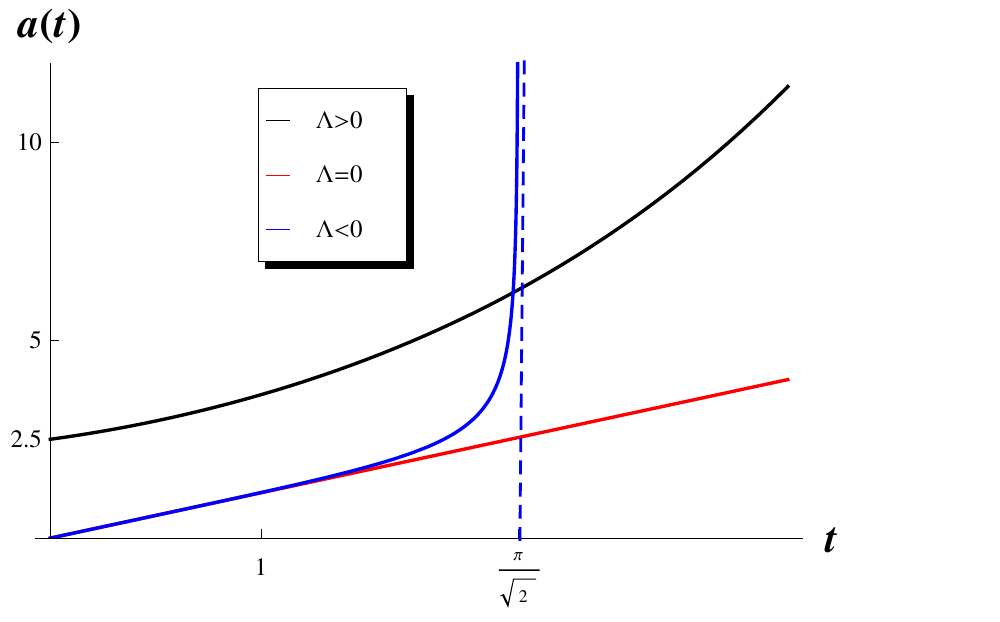}
\includegraphics[width=0.5\textwidth]{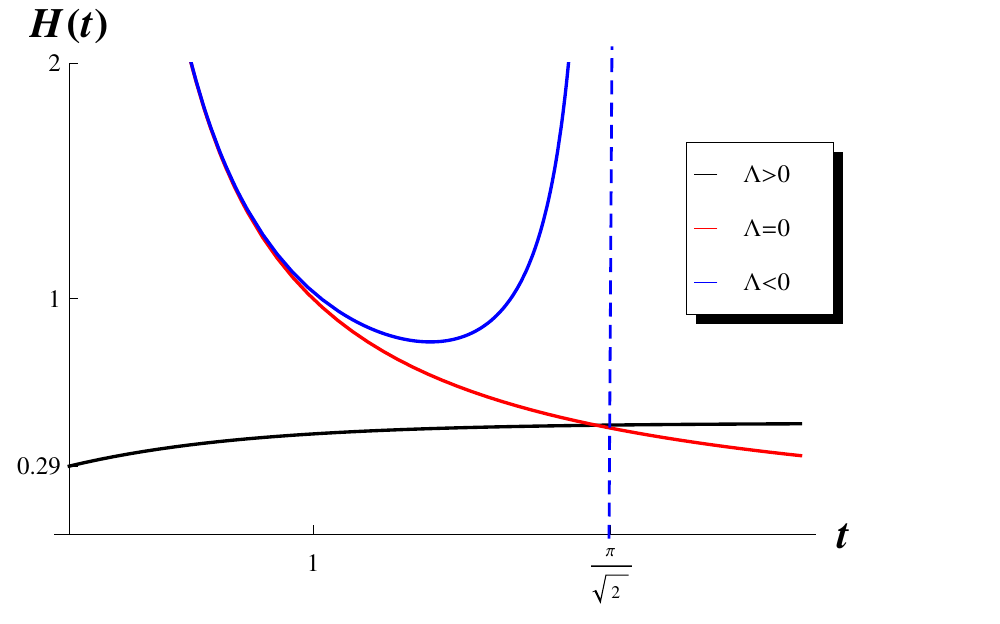}\\
\includegraphics[width=0.5\textwidth]{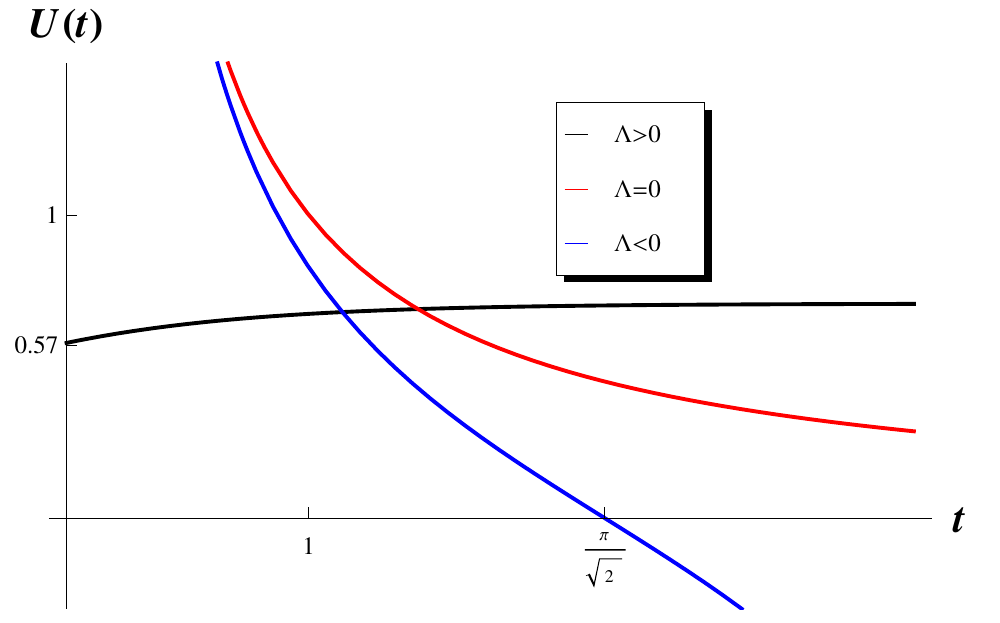}
\includegraphics[width=0.5\textwidth]{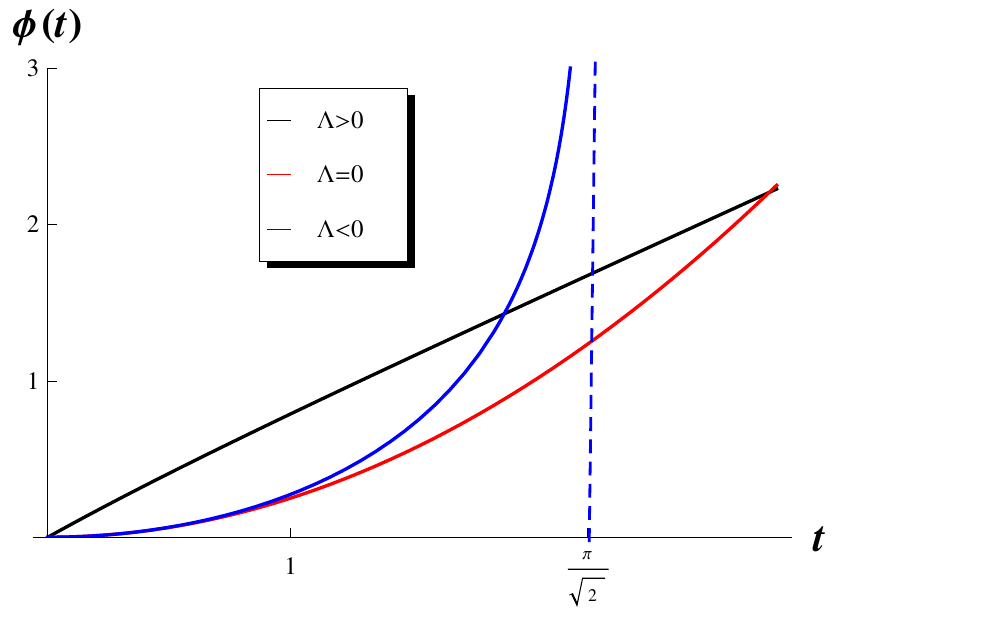}\\
\includegraphics[width=0.5\textwidth]{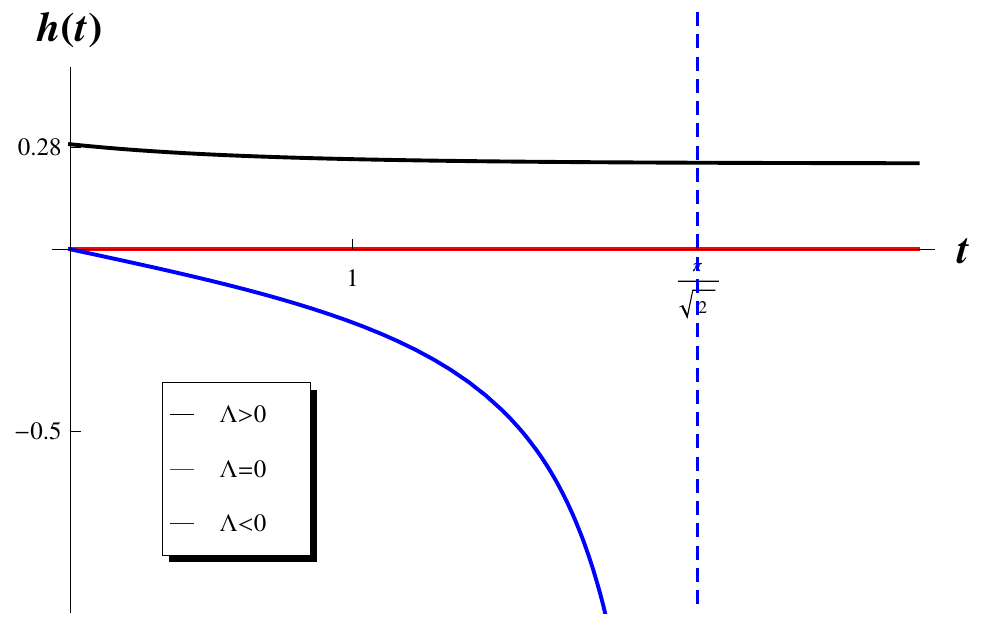}
\includegraphics[width=0.5\textwidth]{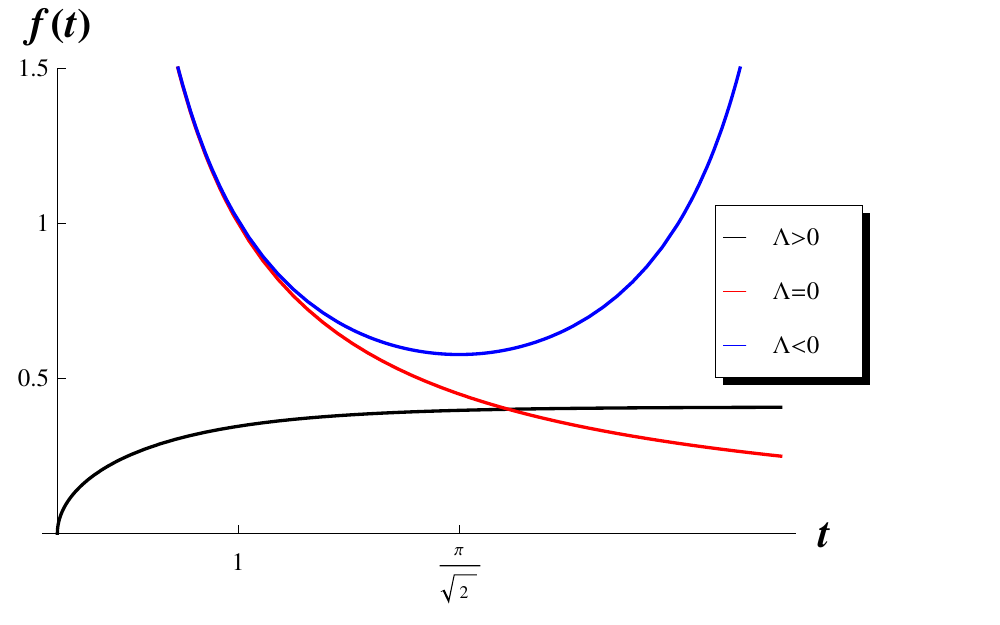}
\caption{For $\Lambda \neq 0$, $t$ and $a(t)$ are measured in units of $|\Lambda|^{-1/2}$, while $U(t)$, $H(t)$, $h(t)$, $f(t)$  and $\phi^{1/2}(t)$ are measured in units of $|\Lambda|^{1/2}$. For $\Lambda = 0$, $a(t)$, $U^{-1}(t)$, $H^{-1}(t)$, $h^{-1}(t)$, $f^{-1}(t)$ and $\phi^{-1/2}(t)$ are measured in the same units of $t$.}
\label{pics}
\end{figure}

In order to solve the system, we first note that equation (\ref{5}) is locally a total derivative (the Euler invariant \cite{Nakahara}). Its integral is
\begin{equation}
\label{6} \frac{U^3}{3} - U f^2 = \frac{C}{a^3} ,
\end{equation}
where $C$ is an integration constant. Assuming $f\neq 0$ in (\ref{4}) implies $\dot{\phi}=(2U)^{-1}$, and replacing (\ref{1}) in (\ref{2}, \ref{3}) allows to integrate $U(t)$ as
\begin{equation}
U(t) = \sqrt{\frac{\Lambda}{2}} \tanh\Big[\sqrt{\Lambda/2} (t+B)\Big],
\end{equation}
for $\Lambda > 0$, where $B$ is an integration constant that sets the origin of time. For $\Lambda < 0$ the solution is
\begin{equation}
U(t) = -\sqrt{\frac{-\Lambda}{2}} \tan\Big[\sqrt{-\Lambda/2} (t+B)\Big],
\end{equation}
and for $\Lambda = 0$
\begin{equation}
U(t) = (t+B)^{-1}.
\end{equation}
Replacing $U$ in (\ref{1}), (\ref{6}) and (\ref{4}), allows to find $f(t)$, $a(t)$ and $h(t)$. The solutions are 
\begin{eqnarray}
a(t) &=& \frac{1}{\sqrt{\Lambda}} \cosh\Big[\sqrt{\Lambda/2} (t+B)\Big] \Bigg(\frac{3 \sqrt{2} C}{\sinh\Big[\sqrt{\Lambda/2} (t+B)\Big]}\Bigg)^{1/3} \\
\phi(t) &=& \frac{1}{\Lambda}\log\Big|\sinh\Big[\sqrt{\Lambda/2}(t+B)\Big]\Big| - \phi_0 \\
h(t) &=& \frac{1}{3} \sqrt{\frac{\Lambda}{2}} \coth\Big[\sqrt{\Lambda/2} (t+B)\Big] \\
f(t) &=& \sqrt{\Lambda\Big(\frac{1}{2} \tanh^2\Big[\sqrt{\Lambda/2} (t+B)\Big]-\frac{1}{3}\Big)}
\end{eqnarray}
for $\Lambda > 0$, and
\begin{eqnarray}
a(t) &=& \frac{1}{\sqrt{-\Lambda}} \cot\Big[\sqrt{-\Lambda/2} (t+B)\Big] \Big(3 \sqrt{2} C \sin^2\Big[\sqrt{-\Lambda/2} (t+B)\Big]\Big)^{1/3} \\
\phi(t) &=& \frac{1}{\Lambda}\log\Big|\sin[\sqrt{-\Lambda/2}(t+B)]\Big|+ \phi_0 \\
h(t) &=& \frac{1}{3} \sqrt{\frac{-\Lambda}{2}} \cot\Big[\sqrt{-\Lambda/2} (t+B)\Big] \\
f(t) &=& \sqrt{-\Lambda\Big(\frac{1}{2} \tan^2\Big[\sqrt{\Lambda/2} (t+B)\Big]+\frac{1}{3}\Big)},
\end{eqnarray}
for $\Lambda < 0$. For $\Lambda=0$,
\begin{eqnarray}
a(t) &=& -(3/2C)^{1/3} (t+B)\\
\phi(t) &=& 1/4 t^2 + B/2 t + \phi_0 \\
h(t) &=& 0\\
f(t) &=& (t+B)^{-1} .
\end{eqnarray}
The integration constant $B$ corresponds to the origin of the time coordinate. For $\Lambda >0$, $B$ can be chosen so that $f(0)=0$, that is $\tanh(\sqrt{\Lambda/2}B)=\sqrt{2/3}$. For $\Lambda \leq 0$, a convenient choice is one for which $a(0)=0$, that is $\sqrt{\Lambda/2} B=\pi/2$. In all cases $\phi_0$ is chosen so that $\phi(0) = 0$ and $C$ is normalized to $1$ and $-1$ for $\Lambda > 0$ and $\Lambda \leq 0$ respectively. With this choice of parameters, the universe expands from a finite size at $t=0$ for $\Lambda>0$, while for $\Lambda\leq 0$ there is an initial singularity.  If $\Lambda < 0$ there is an infinite expansion at a finite cosmic time. These results are depicted in Fig. (\ref{pics}).

\section{Discussion and summary}   
Although probably our model is not very realistic, it shows how a slight modification in the form of the action, which would have negligible effects in a small region around us (e.g., at the scale of the solar system at the present age of the universe), can give rise to dramatic changes in the cosmic evolution. Let us summarize the main features of this very simple scenario:\\
$\bullet$ For $\Lambda > 0$, the solution does not have an initial, point-like singularity. The universe starts with a minimum size and undergoes an accelerated expansion.\\
$\bullet$ For $\Lambda < 0$, there is an initial point-like singularity like in the standard FRWL cosmologies. At late times the universe also experiences an accelerated expansion, reaching an infinity scale factor in finite cosmic time, $t=\pi/\sqrt{-2\Lambda}$.\\
$\bullet$ The time evolution near the big bang is quite sensitive to the sign of $\Lambda$. The solution for $\Lambda = 0$ can not be obtained by taking the limit of $\Lambda \rightarrow 0$ from the $\Lambda \neq 0$ solutions, however its bahaviour is very alike the $\Lambda<0$ case near the big bang.\\
$\bullet$ This expansion does not require dark energy or matter, but is just a dynamical effect of allowing the torsion not to vanish identically. In fact, the ``matter density" (\ref{rho}) and ``pressure" (\ref{p}) are nonzero only because of the nonvanishing torsion.  We find an accelerated expansion for negative cosmological constant. In standard GR an accelerated expansion can be obtained with ordinary matter for $\Lambda>0$, or with $\Lambda<0$ and exotic matter. Here the acceleration is also found for the negative cosmological constant and without matter.\\
$\bullet$ The way we have introduced torsion here is through a slight modification of the standard assumptions of Genral Relativity. The specific nature of the field $\phi$ is not very important, its origin could be exotic or mundane, but the fact that is present makes a huge difference.\\ 
$\bullet$ There are many ways to refine the model by assuming a more realistic, dynamical, scalar field $\phi$. There are other topological invariants that could be included besides the Euler form (Pontryagin, Nieh-Yan); there are couplings between branes of different dimensions and Chern-Simons forms \cite{Miskovic,Edelstein}, that also bring in torsion; finally, the presence of spinning matter also produces some amount of torsion, and although it might be very dilute at the present time, it was certainly not negligible near the big-bang. \\

The present information about the evolution of the universe at large scales provides a reason to revise our cosmological assumptions. The case presented here is just one example of how dramatically different scenarios can be if some of the assumptions one might take for granted --like the absence of torsion--, is removed. There could certainly be others and a critical survey of them would seem most appropriate, rather than sticking to our prejudices and invoking magic to patch up the resulting picture.

\newpage
\textbf{\Large Acknowledgments}\newline
This work was supported by Fondecyt grants \# 1110102, 1100328, 1100755, 1100328, and by Conicyt grant \textit{Southern Theoretical Physics Laboratory, ACT-91}. One of us, Adolfo Toloza, acknowledges financial support by program MECESUP 0605. The Centro de Estudios Cient\'{\i}ficos (CECS) is funded by the Chilean Government through the Centers of Excellence Base Financing Program of Conicyt.

\end{document}